\documentclass[epj]{svjour}
\usepackage{graphics}
\begin{document}

\title{A Unified Framework for the Pareto Law and Matthew Effect using Scale-Free Networks}

\author{Mao-Bin Hu$^1$,\thanks{\emph{E-mail:} humaobin@ustc.edu.cn},
Wen-Xu Wang$^2$,
Rui Jiang $^1$,
Qing-Song Wu$^1$,\thanks{\emph{E-mail:} qswu@ustc.edu.cn}
Bing-Hong Wang$^2$,
Yong-Hong Wu$^3$}

\authorrunning{M.B. Hu, W.X. Wang, Q.S. Wu et al}

\titlerunning {Scale-Free Network Games Provide a Unifying Framework for Pareto Law and Matthew Effect}

\institute{1. School of Engineering Science, University of Science and
Technology of China, Hefei, 230026, P.R.China
\\
2. Nonlinear Science Center and Department of Modern Physics,
University of Science and Technology of China, Hefei, 230026, P.R.China
\\
3. Department of Mathematics and Statistics, Curtin University of Technology,
Perth, WA6845, Australia}

\date{Received: date / Revised version: date}

\abstract{ We investigate the accumulated wealth distribution by 
adopting evolutionary games taking place on scale-free networks.
The system self-organizes to a critical Pareto distribution (1897)
of wealth $P(m)\sim m^{-(v+1)}$ with $1.6 < v <2.0$ (which is in 
agreement with that of U.S. or Japan). 
Particularly, the agent's personal wealth is
proportional to its number of contacts (connectivity), and this
leads to the phenomenon that the rich gets richer and the poor
gets relatively poorer, which is consistent with the Matthew
Effect present in society, economy, science and so on. Though our
model is simple, it provides a good representation of cooperation
and profit accumulation behavior in economy, and it combines the
network theory with econophysics. }

\PACS{
      {87.23.Ge}{Dynamics of social systems} \and
      {89.75.Hc}{Networks and genealogical trees} \and
      {05.10.-a}{Computational methods in statistical physics and nonlinear dynamics}\and
      {89.75.-k}{Complex systems}
}

\maketitle

\section{Introduction}
\label{intro} The interaction of many cooperatively interacting
agents in economy has many features in common with the statistical
physics of interacting systems. A century ago, Pareto (1897)
showed that the probability distribution $P(W)$ for income or
wealth of an individual in the market decreased with the wealth
$W$ according to a power law \cite{Pareto}:
\begin{equation}
P(W) \propto W^{-(1+v)}
\end{equation}
where the value of $v$ was found to lie between 1 and 2
\cite{Dragulescu,Moss,Fujiwara,Levy}.
Studies on real data show that the high-income group indeed
follows the Pareto law, with $v$ varying from 1.6 for USA
\cite{Dragulescu} to 1.8-2.2 in Japan \cite{Moss}.

The previous studies of wealth distribution often adopt an
ideal-gas model in which each agent is represented by a gas
molecule and each trading is a money-conserving collision
\cite{Chakraborti,Dragulescu2,Fischer,Wang,Chatterjee,Xi}. The
model considers a closed economic system where the total money is
conserved and the number of economic agents is fixed. Money and
average money per agent are equivalent to the energy and
temperature in an equilibrium system. Basically, this ideal-gas
model can only reproduce the Gibb distribution or Gaussian-like
stationary distribution of money \cite{Chakraborti}. However,
Chatterjee et al introduce the quenched saving propensity of the
agents, and the system self-organizes to the Pareto distribution
of money with $v \sim 1$ \cite{Chatterjee}. We also note that the
model is not suitable for studying the material wealth
distribution because, in general, the total material wealth of the
system will increase with time \cite{Dragulescu2,Chatterjee}.

The unique feature of our work is that we adopt the scale-free
network to represent the cooperative structure in
population and study the wealth increment by using evolutionary
games as a paradigm for economic activities.

A wide range of systems in nature and society can be described as
complex networks. Since the discovery of small-world phenomena by
Watts and Strogatz \cite{WS} and Scale-free phenomena by
Barab\'{a}si and Albert \cite{BA},investigation of complex
networks has attracted continuous attention from the physics
community \cite{AB}.

Network theory provides a natural framework to describe the
population structure by representing the agents of a given
population with the network vertices, and the contacts between
those agents with edges \cite{Santos}. One can easily conclude
that well-mixed populations can be represented by complete
(fully-connected, regular) networks. Spatially-structured
populations are associated with regular networks, exhibiting a
degree distribution $d(k)$ which is sharply peaked at a single
value of the connectivity $k$, since all agents generally have the
same averaged connectivity. 
Recently, much empirical evidence of real-world social networks 
has revealed that they are associated with a scale-free, 
power-law degree distribution, $d(k)\sim k^{-\gamma}$ with 
$2 \leq \gamma \leq 3$ \cite{AB,Santos,Dorogotsev,Newman}. 
That is, interactions in real-world
networks are heterogeneous that different individuals have
different numbers of average neighbors whom they interact with. 
Thus, the classic regular or random networks are not good 
representations of many real social networks which likely possess the
self-organized mechanism. Hence, in this paper, we adopt the
scale-free network model to construct the cooperation structure in
population.

The evolutionary game theory has been considered to be an
important approach for characterizing and understanding the
cooperative behavior in systems consisting of selfish individuals
\cite{Gintis,Colman}. Since their introduction, the Prisoner's
Dilemma (PD) and the Snowdrift Game (SG) have drawn much attention
from scientific communities \cite{Nowak,Nowak2,Nowak3,Hauert,McNamara}. 
In both games, two players simultaneously decide whether to cooperate
(C) or defect (D). Each player will get a payoff based on his and
his opponent's strategy in each step and then the players will
choose to change their strategy or to keep their strategy
unchanged based on some take-over strategies. One can see that
both games' dynamics are very similar to the cooperation and
payoff activities between agents in economy and so they are
intrinsically suitable for characterizing the payoff and wealth
accumulating behavior in populations.

In this paper, we investigate the wealth accumulation of agents
playing evolutionary games on the scale-free network. The
simulation results show the Pareto wealth distributions along with
some remarkable phenomena including the total wealth variation
with game parameters, and the Matthew Effect in economy, science,
fame, and so on\cite{Merton,Bonitz,Brewer,Wade}.

\section{Model}
\label{model}

In this paper, the simulation starts from establishing the underlying
cooperation network structure according to the most general
Barab\'{a}si-Albert (BA) scale-free network model \cite{BA}.
In this model, starting from $m_0$ fully connected vertices, one vertex
with $m \leq m_0$ edges is attached at each time step in such a way that the
probability $\Pi_i$ of being connected to the existing vertex $i$ is
proportional to the degree $k_i$ of the vertex, i.e. $\Pi_i={k_i \over
\Sigma_j k_j}$, where $j$ runs over all existing vertices.
Initially, an equal percentage of cooperators or defectors was randomly
distributed among the agents (vertices) of the population.
At each time step, the agents play the PD or SG with their neighbours
and get payoff according to the games' payoff matrix.

In the Prisoner's Dilemma, each player can either `cooperate' (invest 
in a common good) or `defect' (exploit the other¡¯s investment).
Two players both receive $R$ upon mutual cooperation and $P$ upon 
mutual defection.
A defector exploiting a cooperator gets an amount $T$ and the
exploited cooperator receives $S$, such that $T>R>P>S$.
So, `defect' is the best response to any action by the opponent
\cite{McNamara}. 
Thus in a single play of the game, each player should defect. 
In the Snowdrift Game (SG), the order of $P$ and $S$ is exchanged,
such that $T>R>S>P$.
Comparing with PD, SG is more in favor of cooperation.
Following common practice \cite{Nowak,Hauert}, we firstly rescale the
games such that each depends on a single parameter.
For the PD, we choose the payoffs to have the values $T=b>1$, $R=1$,
and $P=S=0$, where $1<b \leq 2$ represents the advantage of defectors
over cooperators.
That is, mutual cooperators each gets 1, mutual defectors 0, and D
gets $b$ against C.
The parameter $b$ is the only parameter.
For the SG, we make $T=1+\beta$, $R=1$, $S=1-\beta$, and $P=0$ with
$0<\beta<1$ as the only parameter.

Evolution is carried out by implementing the finite population
analogue of replicator dynamics \cite{Gintis,Hauert}. In each
step, all pairs of directly connected individual $x$ and $y$
engage in a single round of a given game. The total payoff of
agent $i$ for the step is stored as $P_i$. And the accumulative
payoff (Wealth) of agent $i$ since the beginning of simulation is
stored as $W_i$. Then the strategy of each agent (Cooperate or
Defect) is updated in parallel according to the following rule: whenever a
site $x$ is updated, a neighbor $y$ is drawn at random among all
$k_x$ neighbors, and the chosen neighbor takes over site $x$ with
probability:
\begin{equation}
P_{xy}={1 \over 1+ e^{(P_x-P_y)/\gamma}},
\end{equation}
where $\gamma$ characterizes noise introduced to permit irrational choices
\cite{Szabo,Szabo2,Szabo3}, and we make $\gamma=0.1$ as in \cite{Szabo2,Szabo3}.

\section{Simulation Results}
\label{result}

\begin{figure}
\resizebox{1.0\hsize}{!}{\includegraphics{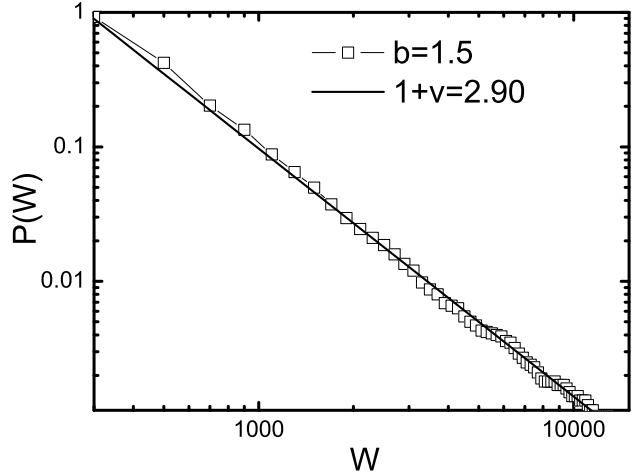}}
\caption{Wealth distribution $P(W)$ for $N=10^4$ agents playing PD
game with $b=1.5$ for $10^5$ steps. The frequency of cooperators
is $0.2137$, and the maximum personal wealth is about 10000.}
\label{fig1}
\end{figure}

\begin{figure}
\resizebox{1.0\hsize}{!}{\includegraphics{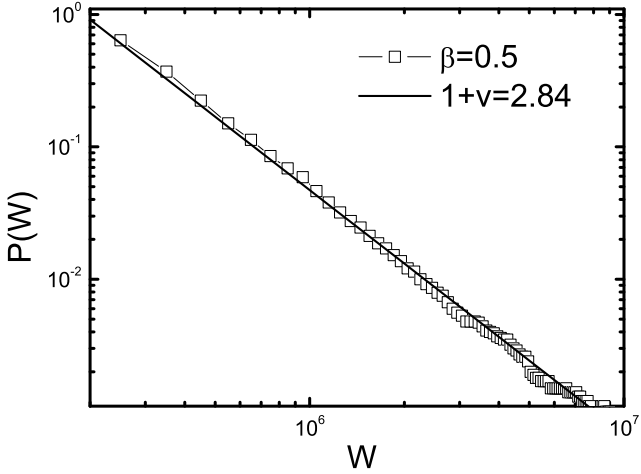}}
\caption{Wealth distribution $P(W)$ for $N=10^4$ 
agents playing SG game with $\beta=0.5$ for $10^5$ steps. 
The frequency of cooperators is $0.9999$, and the maximum personal 
wealth is about $10^7$.} \label{fig2}
\end{figure}

We carry out the simulation for a population of $N=10^4$ agents
occupying the vertices of a BA scale-free network. The
distributions of wealth, total wealth, and k-wealth relation were
obtained after a time period of $T=10^5$ steps.

We first examine the wealth distribution $P(W)$ of the system.
Fig. \ref{fig1} and Fig. \ref{fig2} show the $P(W)$ for PD ($b=1.5$)
and SG ($\beta=0.5$) respectively. One can see that both charts
show power-law distribution of personal wealth which is in
agreement with Pareto's law with $v=1.90$ and $v=1.84$ respectively.
We perform different simulations by altering the values of $b$ and
$\beta$, and the results show similar wealth distributions with
extremely robust power law. For different simulations, the
exponential factor $v$ varies between $1.6$ and $2.0$ that are in
agreement with the empirical values observed in economies
including that of U.S ($1.60$) \cite{Dragulescu} and Japan
($1.80\sim 2.20$) \cite{Moss}. We focus on the payoff and wealth 
accumulating behavior in population. 
In this sense, the wealth distribution we study here essentially
corresponds to `real wealth' or `material wealth', and not the
`paper money' that is generally conserved in the economic system.
We also note that the wealth distribution is independent of the
system size $N$ or the simulation time $T$. Although the system's
maximum personal wealth is different for Fig. \ref{fig1} and
Fig. \ref{fig2} because of the difference in cooperators'
frequency, the power law persists for both high and low
cooperator's frequency cases. All these factors indicate the
robustness of our model to reproduce the Pareto Law of economy.

\begin{figure}
\resizebox{1.0\hsize}{!}{\includegraphics{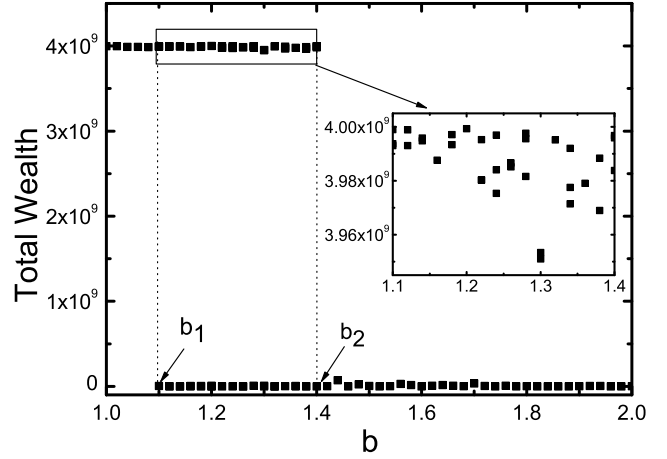}}
\caption{Total Wealth variation for $N=10^4$ agents playing PD game.
The arrows with $b_1=1.10$ and $b_2=1.40$ show the boundaries of 
the bistable region.
The insert shows the fluctuation of the total wealth in the high 
branch of the bistable region.}
\label{fig3}
\end{figure}

\begin{figure}
\resizebox{1.0\hsize}{!}{\includegraphics{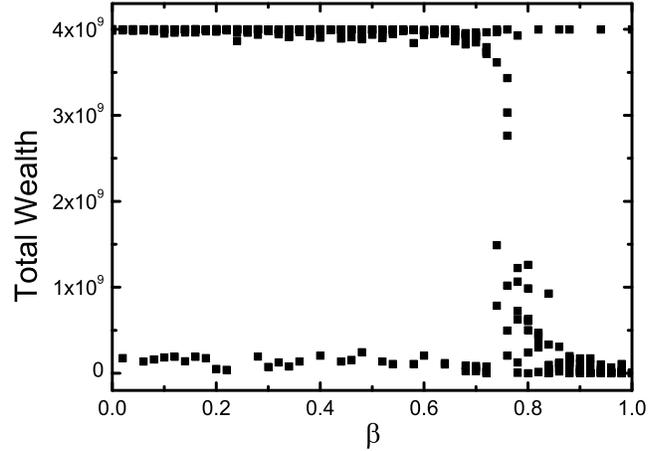}}
\caption{Total Wealth variation for $N=10^4$ agents playing SG game. }
\label{fig4}
\end{figure}

Now we consider the system's total wealth variation with the
parameter $b$ or $\beta$. Fig. \ref{fig3} and Fig. \ref{fig4} show
the variation of total wealth of a $N=10^4$ agents system playing
PD and SG respectively. One can see from Fig. \ref{fig3} that the
total wealth takes a high value ($\approx 4\times 10^9$) when $b$
is relatively small ($\leq 1.10$). Then there is a bistable region
($1.12 < b < 1.40$) where the total wealth can be either high
($\approx 4\times 10^9$) or low ($\approx 5 \times 10^5$). When
$b$ is greater than $1.40$, the total wealth remains low ($\approx
5 \times 10^5$). The high value of the system's total wealth can
be as large as $10^4$ times of the low value. We note that the
total wealth value is related to the frequency of cooperators such
that the system's total wealth is high when the frequency is high,
and a low total wealth shows up when the frequency is low. For
instance, the frequency of cooperators is $0.9999$ and the maximum
total wealth is 3996720318 when $b=1.0$. However, the frequency of
cooperators is only $0.2137$ and the total wealth is only 5461747
when $b=1.5$. This phenomenon implies that when the advantage of
defectors over cooperators is too high, the system will take the
risk of sharply reducing its total wealth. Thus, a
defector-favored economic rule can prohibit the emergence of
cooperators and, what is more, greatly reduce the total wealth of
the system.

However, because the SG payoff matrix $T>R>S>P$ is intrinsically
cooperator-favored, the total wealth for SG fluctuates as the
$\beta$ value changes as shown in Fig. \ref{fig4}.

\begin{figure}
\resizebox{1.0\hsize}{!}{\includegraphics{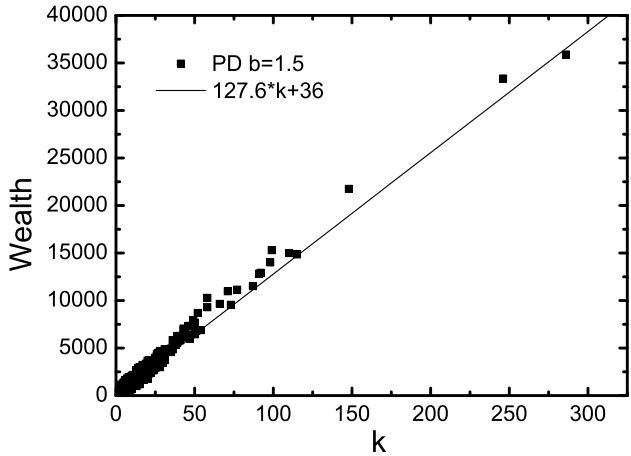}}
\caption{K-Wealth relation for $N=10^4$ agents playing PD game
with $b=1.5$.}
\label{fig5}
\end{figure}

\begin{figure}
\resizebox{1.0\hsize}{!}{\includegraphics{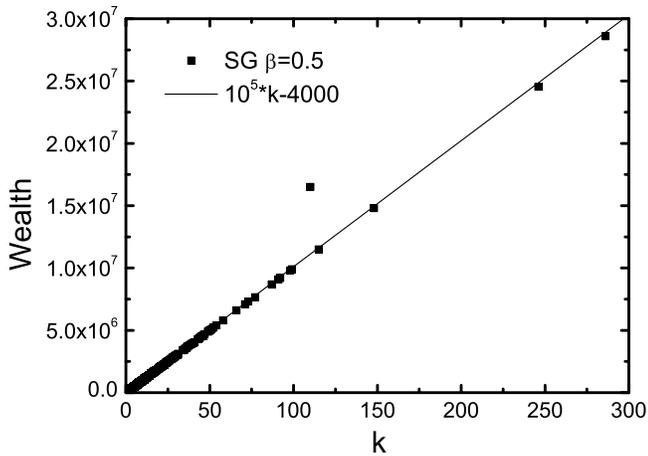}}
\caption{K-Wealth relation for $N=10^4$ agents playing SG game
with $\beta=0.5$.}
\label{fig6}
\end{figure}

Fig. \ref{fig5} and Fig. \ref{fig6} show the relation of personal
wealth $W$ with its connectivity $k$. One can see in both cases
(PD and SG) that the personal wealth is proportional to its
connectivity.
Since the number of agents it contacts reflects the information
resources it has, this model also provides a framework to explain
the fact that agents with more information resources can gain more
profit in modern society's economy.

\begin{figure}
\resizebox{1.0\hsize}{!}{\includegraphics{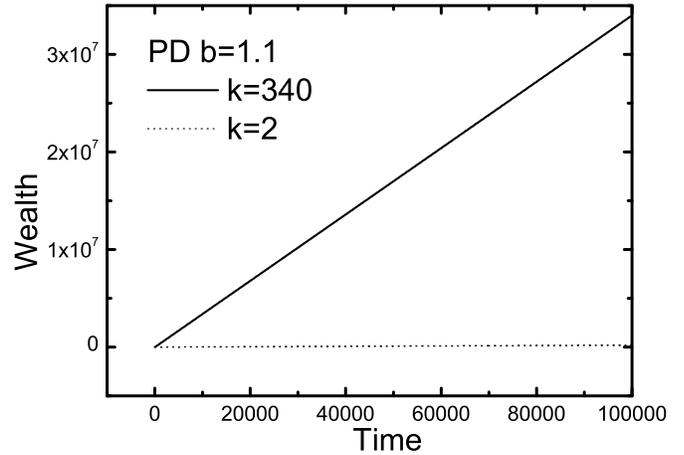}}
\caption{Matthew Effect in PD game. The one with
more connectivity surpass the one with fewer connectivity in their
personal wealth. } \label{fig7}
\end{figure}

\begin{figure}
\resizebox{1.0\hsize}{!}{\includegraphics{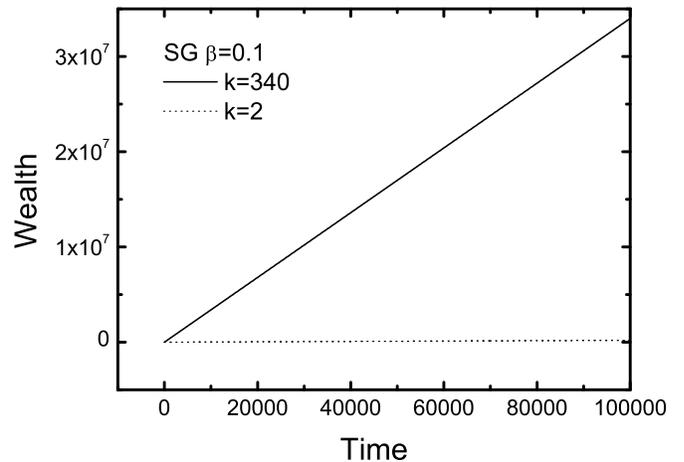}}
\caption{Matthew Effect in SG game.}
\label{fig8}
\end{figure}

This proportional relation between personal wealth and its
connectivity is also a possible mechanism for the emergence
of the Matthew Effect in economy. The ``Matthew Effect" refers to
the idea that in some areas of life (wealth, achievement, fame,
success et al), the rich gets richer and the poor gets poorer
\cite{Merton,Bonitz,Brewer,Wade}. The eminent sociologist Robert
Merton used the term ``Matthew effect" to describe the deplorable
practice of scientists giving exclusive credit to the most
distinguished one among several equally deserving candidates
\cite{Merton}. The Matthew effect for Countries (MEC) was also
discovered \cite{Bonitz}. Our simulations capture a possible
underlying mechanism for these phenomena. In Fig. \ref{fig7} and
Fig. \ref{fig8}, the wealth variations of two individual agents are
compared. One can see that with both PD and SG, the wealth of the
agent with more connectivity exceeds the agent with less
connectivity.
We note that this tendency remains the same when different values
of parameter $b$ or $\beta$ are used. And also the tendency is
independent of the system size $N$ or the simulation time $T$.
Thus, the agents with more cooperation partners will get richer
and richer while those with fewer partners will get relatively
poorer. It is true, from our experience, that a successful people
(company, country etc) usually have more partners than a
unsuccessful one, and this huge relation network will provide him
more profits. So, to some extent, our model explains the Matthew
Effect from a statistical point of view.

\section{Conclusions}
\label{conclusion}

In conclusion, we have studied the wealth distribution in economy
by calculating the accumulative payoff of agents involving in
revolutionary games on the cooperation network with scale-free
property. The simulations confirm Pareto's power law of wealth
distribution. And the values of exponential factor $v$ are in
agreement with the empirical observations.

The simulation shows that the system's total wealth varies with
the game parameters. The results of the PD game shows that agents
tend to cooperate with a frequency of nearly $1.0$ and a high
total wealth can be achieved when the advantage of defector over
cooperator ($b$) is relatively low. But the total wealth will drop
to a very low value when $b$ is high.
The total wealth of SG fluctuates as the $\beta$ value changes.

The model also provides a possible explanation for the Matthew
Effect from a statistical physics point of view. The simulations
show that the agents' personal wealth is proportional to the
number of its contacts (connectivity).
This leads to the phenomenon that the rich gets 
richer and the poor gets poorer (Matthew Effect).
Thus, in this sense, one has to increase the number of partners
in order to gain more profit in modern society.
This also suggests a framework to explain why agents with more
information resources can gain more profit in modern society's
economy, since the connectivity is a representation of an agent's
information resource.

It is evident from the above discussions that, our model provides
a simple but good approach to study the wealth phenomena in
economy, and therefore is worthy of more attention.

\section*{ACKNOWLEDGEMENTS}
\label{Thanks}
This work is financially supported by the National Natural Science
Foundation of China (Grant No. 10532060, 10404025) and the Australian
Research Council through a Discovery Project Grant.

\end{document}